\documentclass[preprint,nofootinbib,longbibliography,
amsmath,amssymb,aps,prd]{revtex4-2}

\usepackage[utf8]{inputenc}
\usepackage{amsmath}
\usepackage{amsfonts,dsfont}
\usepackage{mathrsfs}
\usepackage{cancel} 
\usepackage{amssymb,ulem}
\usepackage{amsmath}
\usepackage{subfigure}
\usepackage{graphicx}% Include figure files
\usepackage{bm}% bold math
\usepackage{enumerate}
\usepackage[colorlinks=true,linkcolor=blue,urlcolor=blue,filecolor=black,
citecolor=red,pdfstartview=FitV,pdftitle={},pdfsubject={},pdfkeywords={},
pdfpagemode=None,bookmarksopen=true]{hyperref}
 \usepackage{appendix}

\usepackage{xcolor}
\usepackage{float}
\usepackage{enumitem}  
\usepackage{bbold}
\usepackage{tikz}
\usepackage{multirow}
\usepackage{longtable}
\usepackage{supertabular,rotating}
\newcommand{\dbar}{\overline{d}}
\newcommand{\ubar}{\overline{u}}

\begin{document}
	\preprint{APS/PRD}
\title{\boldmath
Light baryon static properties in dispersive approach}
	
	\author{Shuge Zeng$^a$}\email{sgzeng@stu2021.jnu.edu.cn}
	\author{Hsiang-nan Li$^b$}\email{hnli@phys.sinica.edu.tw}
\author{Fanrong Xu$^a$}
	\email{fanrongxu@jnu.edu.cn}
	
\affiliation{$^a$Department of Physics, College of Physics and Optoelectronic Engineering, 
Jinan University, Guangzhou 510632, P.R. China}
\affiliation{$^b$Institute of Physics, Academia Sinica,
Taipei, Taiwan 115, Republic of China}
	
	\begin{abstract} 
We extend our dispersive analyses on meson static properties to those of light baryons. 
The formalism treats the dispersion relation, which a baryonic correlation function 
obeys, as an inverse problem, solve for the involved spectral density with available
operator-product-expansion (OPE) inputs directly, and extract baryon 
static properties from the spectral density. We observe that the simultaneous 
implementation of the chiral-even and chiral-odd dispersive constraints unambiguously 
determines baryon masses and pole residues. A common set of quark and gluon condensates, 
which appear in OPE factorization and are universal, is found to accommodate the masses 
of a $\rho$ meson, a proton and a $\Delta(1232)$ baryon. The advantage of our 
approach over the conventional handling of QCD sum rules is advocated. This work 
encourages broad applications of our nonperturbative analytical method to baryon systems.

	\end{abstract}
	\maketitle
	\clearpage
	% \tableofcontents
	\clearpage
	\newpage
	\pagenumbering{arabic}
%%%%%%%%%%%%%%%%%%%%%%%%%%%%%%%%%%%%%%%%%%%%%%%%%%%%%%%%%%%%%%%
\section{INTRODUCTION}

It has been recognized that QCD sum rules \cite{SVZ}, as a popular nonperturbative 
method, suffer significant theoretical uncertainties from the assumption of quark-hadron 
duality for parametrizing a spectral density and from the discretionary prescription for 
designating a stability window in a Borel mass \cite{Coriano:1993yx,Coriano:1998ge,
Leinweber:1995fn,Huang:1998wj,Harnett:2000fy,Gubler:2010cf}. To remedy the above drawbacks, 
we proposed to handle QCD sum rules as an inverse problem \cite{Li:2020ejs}; the spectral 
density on the hadron side of a dispersion relation, including both resonance and continuum 
contributions, is regarded as an unknown in an integral equation, which is solved with 
standard operator-product-expansion (OPE) inputs on the quark side. Once a dispersion 
relation is solved directly, there is no need to presume the existence of a resonance, 
to introduce a free continuum threshold, to implement the quark-hadron duality, or to
assign discretionary weights to continuum contributions and power corrections in an OPE.
In other words, our approach relies solely on analyticity of physical observables and 
positivity of spectral densities. Moreover, the precision of predictions can be improved 
systematically by adding subleading contributions to an OPE \cite{Li:2021gsx}.

Our formalism is similar to the one developed in \cite{Karateev:2019ymz,He:2023lyy,He:2025gws}, 
which also inputs high-energy OPEs with quark and gluon condensates to bootstrap 
low-energy dynamics. The idea behind both frameworks stems from the $S$-matrix bootstrap 
conjecture postulated by Geoffrey Chew in 1960s \cite{Chew:1962mpd}; a well-defined infinite 
set of self-consistency conditions (analyticity, unitarity, crossing, etc.) determines 
uniquely the aspects of particles in nature \cite{Cushing:1985zz,vanLeeuwen:2024uzj}. 
The difference between our approach and \cite{Karateev:2019ymz,He:2023lyy,He:2025gws}
to, say, a two-current correlator, is that constraints on other observables like form 
factors, in addition to those on a spectral density, were incorporated in the latter. The 
inclusion of form factors is based on the assumption that the spectral density 
associated with the correlator is saturated by two pion states in an intermediate energy 
region. However, Refs. \cite{Karateev:2019ymz,He:2023lyy,He:2025gws} considered only the 
constraints imposed by the first few moments of a spectral density through sum rules, while 
we concentrate on a spectral density and attempt to retrieve full information from it. We 
mention an alternative application of analyticity \cite{Pelaez:2025jrn}, where various 
meson resonance poles were extracted from dispersion relations with the inputs of 
$\pi\pi$ scattering data.

The dispersive analyses on nonperturbative quantities of several mesonic states have been 
conducted, covering masses and decay constants \cite{Li:2020ejs,Li:2021gsx,Zhao:2024drr,
Mutuk:2024jvv,Li:2024fko} for series of $\rho$ resonances and glueballs, and the pion 
light-cone distribution amplitude \cite{Li:2022qul}. It was extended to investigations of 
hadronic properties in nuclear matter recently \cite{Mutuk:2025lak}. Besides, the perspective 
of treating a dispersion relation as an inverse problem was generalized to the understanding 
of neutral meson mixing parameters \cite{Li:2020xrz,Xiong:2022uwj,Li:2022jxc} and to the 
estimation of the hadronic vacuum-polarization contribution to the muon anomalous magnetic 
moment \cite{Li:2020fiz}. Here we will further extend the formalism for meson static 
properties to those of light baryons. A challenge arises immediately from the splitting of 
a correlation function defined by baryonic currents into a chiral-even part and a 
chiral-odd part. The OPE for the former (latter) contains only dimension-even (dimension-odd) 
condensates. It has been observed \cite{Li:2020ejs} that a sufficient number of power 
corrections are necessary for establishing a physical resonance solution. After the condensates 
of different dimensions are separated, the number of power corrections associated with each 
chirality is not enough for identifying a physical solution.

The difficulty can be overcome by the strategy below. We gather baryon masses and pole 
residues allowed by each dispersion relation, and find the common ones between the two sets 
of solutions. It turns out that baryon masses and pole residues can be specified 
unambiguously in this way. The reason for the success of the strategy is that the 
simultaneous implementation of the chiral-even and chiral-odd dispersive constraints takes 
into account all available power corrections in some sense. To highlight the strength of our 
approach, we explicate that no stability window in a Borel mass exists in the conventional 
handling of QCD sum rules for the proton mass, so a definite determination is unlikely. 
Since the OPE formulation abides by short-distance factorization, involved condensates 
of quark and gluon fields are universal. We demonstrate that the same set of quark and gluon 
condensates can account for the masses of a $\rho$ meson, a proton and a $\Delta(1232)$ 
baryon collected in \cite{PDG}. This work encourages intensive applications of our 
nonperturbative analytical method to baryon systems.

The rest of the paper is organized as follows. We recapture the derivation of the 
$\rho$ meson mass and decay constant in Sec.~II, focusing on the inverse matrix method for
solving a dispersion relation \cite{Li:2021gsx}. The framework is applied to the proton and 
$\Delta(1232)$ baryon cases in Sec.~III. The aforementioned strategy to establish physical 
solutions from chiral-even and chiral-odd dispersion relations is elaborated. The advantage 
of our approach over the conventional QCD sum rules is emphasized. We then conclude in 
Sec.~IV.

\section{INVERSE MATRIX METHOD}

The key ingredients of our theoretical setup have been elucidated in \cite{Li:2021gsx},
including the construction of the dispersion relation from a correlation function,
which connects the dispersive integral of a spectral density to the corresponding OPE.
Taking the simple $\rho$ meson case as an example, we explain how to deduce the $\rho$ 
meson mass and decay constant from the dispersion relation. The two-point correlator 
\begin{equation}
\Pi^{\mu\nu}(q^2)=\int d^4 x e^{i q\cdot x}\langle 0| T\left[ J^{\dagger \mu}(x) J^\nu(0)
\right]| 0\rangle
%=&\int d^4 x e^{i q\cdot x}\Pi^{\mu\nu}(x)\nonumber\\
= (q^\mu q^\nu -q^2 g^{\mu\nu}) \Pi(q^2),
\label{eq:tpc}
\end{equation}
defines the the vacuum polarization function $\Pi(q^2)$, where $T$ denotes the time-ordered 
product, and the momentum $q$ is injected into the interpolating vector current 
$J^\mu=(\ubar\gamma^\mu u -\dbar \gamma^\mu d)/\sqrt{2}$. The OPE for $\Pi(q^2)$ is reliable 
in the deep Euclidean region of $q^2$, and available up to the dimension-six condensate, i.e., 
to the power correction of $1/(q^2)^3$ as shown below \cite{SVZ}. 

The analyticity of the function $\Pi(q^2)$ leads to the dispersion relation \cite{Li:2021gsx}
\begin{equation}
    \int_0^R ds\frac{\rho(s)}{s-q^2}=\frac{1}{\pi}\int_0^R ds \frac{\text{Im}\Pi^{\text{pert}}(s)}{s-q^2}+\frac{1}{12\pi}\frac{\langle\alpha_sG^2\rangle}{(q^2)^2}+\frac{2\langle m_q\bar{q}q\rangle}{(q^2)^2}+\frac{224}{81}\frac{\kappa\alpha_s\langle\bar{q}q\rangle^2}{(q^2)^3},
    \label{sumrulerho}
\end{equation}
where $\rho(s)\equiv \text{Im}\Pi(s)/\pi$ is the spectral density, 
$\text{Im}\Pi^{\text{pert}}(s)=(1+\alpha_s/\pi)/(4\pi)\equiv c\pi$ is the imaginary part of 
the perturbative piece in the OPE, and $R$ is an arbitrary ultraviolet cutoff of the 
integration variable $s$. Precisely speaking, $R$ is the radius of the big circle as part of
the contour on a complex plane, along which the integration of $\Pi(s)$ is performed. In 
Eq.~(\ref{sumrulerho}) 
$\langle\alpha_s G^2\rangle\equiv \langle \alpha_s G^a_{\mu\nu}G^{a\mu\nu}\rangle$ 
is the gluon condensate, $m_q$ is a light quark mass, and the parameter $\kappa=2$-4 
\cite{Chung:1984gr,Narison:1995jr,Narison:2009vy} quantifies the violation in the 
factorization of the four-quark condensate $\langle (\bar q q)^2\rangle$ into the square 
of the quark condensate $\langle \bar q q\rangle$.

Working on the subtracted spectral density 
\begin{equation}
    \Delta\rho(s,\Lambda)=\rho(s)-\frac{1}{\pi}\text{Im}\Pi^{\text{pert}}(s)[1-\exp(-s/\Lambda)],
    \label{ssd}
\end{equation}
we can push $R$ toward the infinity, arriving at
\begin{equation}
    \int_0^\infty dy\frac{\Delta\rho(y)}{x-y}=\int_0^\infty dy \frac{ce^{-y}}{x-y}
    -\frac{1}{12\pi}\frac{\langle\alpha_sG^2\rangle}{x^2\Lambda^2}
    -\frac{2\langle m_q\bar{q}q\rangle}{x^2\Lambda^2}
    -\frac{224}{81}\frac{\kappa\alpha_s\langle\bar{q}q\rangle^2}{x^3\Lambda^3},
    \label{sumrulerho_changed}
\end{equation}
where the variable changes $x=q^2/\Lambda$ and $y=s/\Lambda$ have been exerted. Because 
$\Delta\rho(s,\Lambda)$ is a dimensionless quantity, it has been cast into the form 
$\Delta\rho(y=s/\Lambda)$. It is seen that the arbitrary cutoff $R$ is replaced by the 
arbitrary scale $\Lambda$, which moves into the power corrections after the variable changes. 
The subtraction function $1-\exp(-s/\Lambda)$ decreases like $s$ at small $s$, and approaches 
unity at large $s\gg\Lambda$, such that $\Delta\rho(s,\Lambda)$ retains the behavior of 
$\rho(s)\sim s$ near the threshold $s\to 0$ \cite{Kwon:2008vq}, and diminishes quickly as 
$s>\Lambda$. We have confirmed that other subtraction functions with the similar boundary and 
asymptotic behaviors yield basically identical solutions for $\rho(s)$. The arbitrariness of
$\Lambda$, to which a physical observable should be insensitive, will help identify the physical 
solution to the dispersion relation. 

%Note that the quark-hadron duality for the unknown 
%spectral density is not assumed at any finite $s$ in the construction.

Equation~(\ref{sumrulerho_changed}) is classified as the first kind of Fredholm integral 
equations in the typical expression 
\begin{equation}
    \int_0^\infty dy\frac{\Delta\rho(y)}{x-y}=\omega(x),
    \label{eq:integral function1}
\end{equation}
where the unknown function $\Delta\rho(y)$ will be solved with the given input function 
$\omega(x)$. Techniques are available for solving the above potentially ill-posed inverse 
problem. Here we adopt the inverse matrix method proposed in \cite{Li:2021gsx}. If 
$\Delta\rho(y)$ decreases rapidly with the variable $y$, the kernel $1/(x-y)$ in 
Eq.~(\ref{eq:integral function1}) can be expanded into a power series of $1/x$. This 
argument also holds for the dispersive integral on the right-hand side of 
Eq.~(\ref{eq:integral function1}), for the numerator $ce^{-y}$ descends fast in 
$y$. The function $\omega(x)$, representing the OPE on the right-hand side of 
Eq.~(\ref{sumrulerho_changed}), can be expanded into a power series of $1/x$ too. Hence, 
we write
\begin{equation}
    \frac{1}{x-y}=\sum^N_{m=1}\frac{y^{m-1}}{x^m},\quad\omega(x)=\sum_{n=1}^N\frac{b_n}{x^n},
    \label{eq:expansion}
\end{equation}
up to some integer $N$. The two sides of Eq.~(\ref{eq:integral function1}) 
can then be matched by equating the coefficients at each power of $1/x$.  

To facilitate the matching, we reformulate Eq.~(\ref{eq:integral function1}) into a matrix form 
by expanding $\Delta\rho(y)$ in terms of a complete set of orthogonal polynomials. As discussed 
in \cite{Li:2021gsx}, the Laguerre polynomials in the support $[0,+\infty)$, with a weight 
factor $e^{-y}$ that suppresses the contribution from large $y$, serve the purpose. The expansion
\begin{equation}
    \Delta\rho(y)=\sum_{n=1}^Na_ny^\alpha e^{-y}L_{n-1}^{(\alpha)}(y),
    \label{eq:LGP}
\end{equation}
and Eq.~(\ref{eq:expansion}) define the matrix elements
\begin{equation}
    M_{nm}=\int_0^\infty dyy^{m-1+\alpha}e^{-y}L_{n-1}^{(\alpha)}(y), 
    \label{eq:Matrix}
\end{equation}
with $m$ and $n$ running from $1$ to $N$, and transform Eq.~(\ref{eq:integral function1}) into 
$Ma=b$. The vector $a=(a_1,a_2,...,a_N)$ groups the unknown coefficients in Eq.~(\ref{eq:LGP}), 
and the vector $b=(b_1,b_2,...,b_N)$ comes from the known input $\omega(x)$ in 
Eq.~(\ref{eq:expansion}). The integer $N$ will be fixed later, and the choice of the 
index $\alpha$ depends on the behavior of $\Delta\rho(y)$ adjacent to the boundary $y= 0$. 
Simply inversing the matrix equation, we obtain the vector $a=M^{-1}b$, which constructs the 
function $\Delta\rho(y)$ according to Eq.~(\ref{eq:LGP}).

We solve for the unknown $a=M^{-1}b$ with the input $b$ from Eq.~(\ref{sumrulerho_changed})
in the inverse matrix method, the subtracted spectral density 
$\Delta\rho(s,\Lambda)$ and then the spectral density $\rho(s)$ in Eq.~(\ref{ssd}). In 
principle, the true solution is approached by increasing the number $N$ of the Laguerre 
polynomials in Eq.~(\ref{eq:LGP}). Nevertheless, both $m$ and $n$ have to be terminated at 
a finite value in a practical operation, because the determinant of the matrix $M$ diminishes 
with its dimension. If $N$ keeps growing, a tiny fluctuation of the input $b$ will be 
amplified by the huge elements of the inverse matrix $M^{-1}$, and positivity of a spectral 
density is lost. This is a generic feature of an ill-posed inverse problem. Therefore, the 
optimal choice of $N$ is set to the maximal integer, around which a solution is stable 
against the variation of $N$. We take the input
parameters
\begin{equation}
    \begin{split}
        &\langle\alpha_sG^2\rangle=0.028\;\text{GeV}^4,\quad\langle 
        m_q\bar{q}q\rangle=0.007\times(-0.215)^3\;\text{GeV}^4,\\
        &\alpha_s\langle\bar{q}q\rangle^2=6.64\times10^{-5}\;\text{GeV}^6\,\quad
        \alpha_s=0.5,\quad\kappa=2,
    \end{split}\label{cond}
\end{equation}
for a numerical illustration. These values are similar to those in \cite{Li:2021gsx},
except the gluon condensate $\langle\alpha_sG^2\rangle$, which is much lower than 
0.08 GeV$^4$ in \cite{Li:2021gsx}. We point out that there exists a range for this
condensate \cite{Eidelman:1978xy,Launer:1983ib,Novikov:1983jt,Yndurain:1999pb} as summarized 
in Eq.~(11) of \cite{Baldo:2003id}. The choice of the gluon condensate in Eq.~(\ref{cond}) 
is still above the lower bound $0.01$ GeV$^4$, while the choice $0.08$ GeV$^4$ is close to 
the upper bound. The adjustment of the parameters aims at a simultaneous accommodation 
of $\rho$ meson, proton and $\Delta(1232)$ static properties in this work.

\begin{figure}
    \centering
    \includegraphics[width=0.75\linewidth]{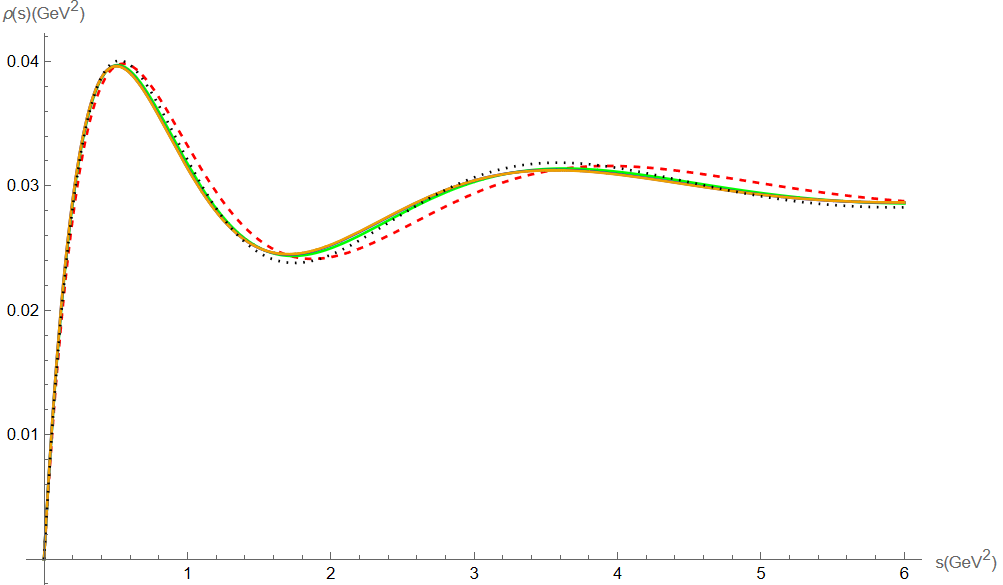} 
    \caption{Dependencies of $\rho(s)$ on $s$ for $\Lambda = 2\; \text{GeV}^2$ and 
    $N=13$ (dashed line), $N=15,16,17$ (solid lines) and $N=19$ (dotted line).}
    \label{fig:2}
\end{figure}

Figure~\ref{fig:2} displays the curves of the spectral density $\rho(s)$ for several $N$'s 
with $\Lambda=2$ GeV$^2$. It is obvious that the curves labeled by $N=15$, 16 and 
$17$ are pretty similar, exhibiting the stability in $N$. To substantiate such stability, we 
resort to a simplified version of the Frechet distance between two curves corresponding to 
adjacent $N$'s (with the same $\Lambda$),
\begin{equation}
    F_{\Lambda,i}[\rho(s)]=\max\{\text{D}\left(\rho_i(s),\;
    \rho_{i+1}(s)\right)\}|_{s\in[0,6\;{\rm GeV}^2]}.\label{fre}
\end{equation}
In the above definition the argument $s$ takes a value in the interval $[0,6\;{\rm GeV}^2]$, 
and $\text{D}\left(\rho_{i}(s),\;\rho_{i+1}(s)\right)$ is the distance between the points on 
the two curves associated with $N=i$ and $N=i+1$. For two identical curves, the Frechet 
distance vanishes; for two distinct curves, the smaller the Frechet distance is, the more 
similar the two curves are. We attain the minimal Frechet distance 
$F_{\Lambda=2\;\text{GeV}^2,i=16}$ by scanning the range $[12,20]$ of $N$, which 
suggests that $N=16$ is the best choice to stabilize $\rho(s)$ in $N$ for 
$\Lambda=2$ GeV$^2$. The best $N$'s for other $\Lambda$ values can be acquired in the same 
manner. The first (major) peak of $\rho(s)$ for $\Lambda=2$ GeV$^2$ located at 
$s\approx 0.60$ GeV$^2$ in Fig.~\ref{fig:2} coincides with the $\rho(770)$ meson mass. The 
oscillatory tail at $s>2$ GeV$^2$ manifests that the continuum contribution to $\rho(s)$ 
deviates from the perturbative one $c= 0.029$. Namely, the local quark-hadron duality 
does not hold actually.

\begin{figure}[h]
    \centering
    \includegraphics[width=0.75\linewidth]{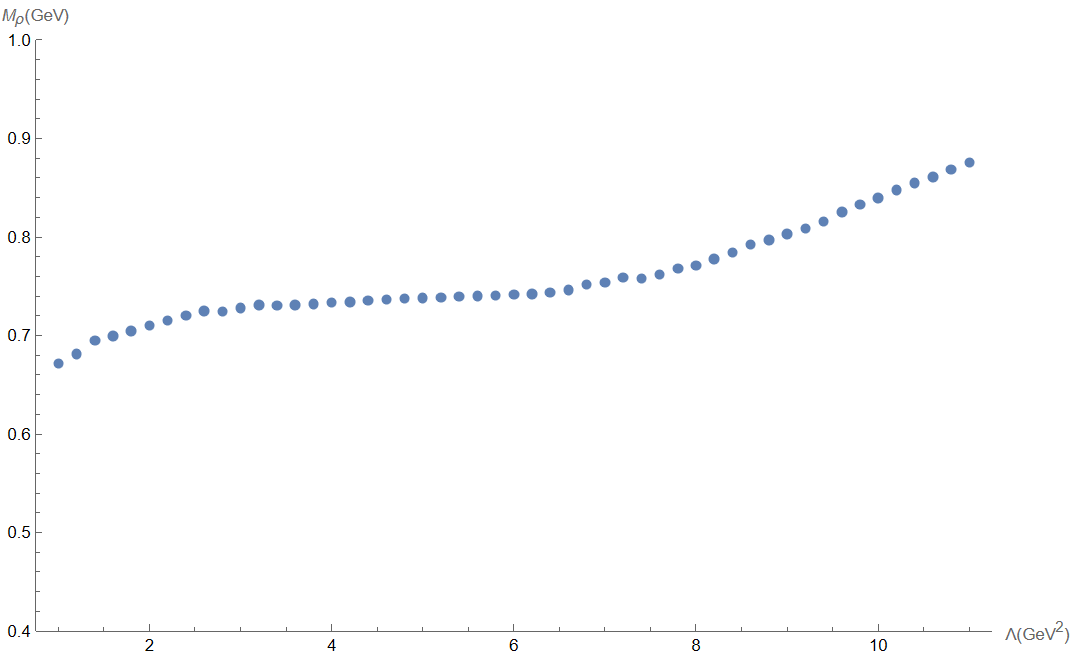}
    \caption{Dependence of the $\rho$ meson mass $M_\rho$ on $\Lambda$. The specific $N$ 
    is taken for each $\Lambda$ that minimizes the Frechet distance in Eq.~(\ref{fre}).}
    \label{fig:3}
\end{figure}

As stressed before, a physical observable should be insensitive to the variation of the
arbitrary scale $\Lambda$. Equation~(\ref{sumrulerho_changed}) shows that the condensate 
effects decrease with $\Lambda$, and diminish in the limit $\Lambda\to\infty$. It implies 
that $\rho(s)$ depends only on the scaling variable $y=s/\Lambda$ at large $\Lambda$, so a 
peak of $\rho(s)$ would drift in $s$ with $\Lambda$. Then the peak location, failing the 
stability criterion, cannot be interpreted as a physical state. On the other hand, $\Lambda$ 
cannot be too small, otherwise the enhanced condensate contributions break the OPE. A 
proper window in $\Lambda$ is expected to exist, within which the OPE, i.e., the reliability 
of solutions, is justified, and the stability of the $\rho$ meson mass sustains. 
Figure~\ref{fig:3}, depicting the $\Lambda$ dependence of the $\rho$ meson mass $M_\rho$, 
reflects the aforementioned features; the curve climbs from $\Lambda=1.0$ GeV$^2$, goes up 
and down mildly about $M_\rho=0.74$ GeV for 3.0 GeV$^2\le\Lambda \le 7.0$ GeV$^2$, and 
then ascends monotonically with $\Lambda$ as $\Lambda> 8.0$ GeV$^2$. We read off the $\rho$ 
meson mass from the flat segment of the curve between $\Lambda=4\;\text{GeV}^2$ and 
$\Lambda=6\;\text{GeV}^2$ in Figure~\ref{fig:3}, 
\begin{equation}
M_\rho=(0.737\pm0.004)\;\text{GeV},\quad\label{mrho}
 \end{equation}
whose uncertainty evinces the salient stability of our result in $\Lambda$. The $\rho$ meson 
mass depends weakly on the OPE parameters \cite{Li:2021gsx}; the representative variation of 
the gluon condensate $\langle\alpha_sG^2\rangle$ by $\pm 20\%$ changes $M_\rho$ by $\mp 5\%$, 
which is not included in Eq.~(\ref{mrho}). The above outcome is close to $(0.78\pm 0.03)$ GeV 
in \cite{Li:2021gsx}, though the inputted gluon condensate in Eq.~(\ref{cond}) is lower; the 
reduction of the gluon condensate has been more or less compensated by that of the quark 
condensate in the present analysis.

At last, we estimate the $\rho$ meson decay constant following \cite{Li:2021gsx}
\begin{equation}
    f_\rho^2\approx\int_0^{a}ds\Delta\rho(s,\Lambda),\label{dec}
\end{equation}
where the upper bound $a$ of the integration variable $s$ is set to the location of the 
first valley in $\Delta\rho(s,\Lambda)$. Compared to \cite{Li:2021gsx} which set 
$a=\infty$, Eq.~(\ref{dec}) excludes potential minor contributions from excited states 
mostly. We get, for $\Lambda\in [4\; {\rm GeV}^2,6\; {\rm GeV}^2]$, the decay constant  
$f_\rho\approx 0.218$ GeV, consistent with 0.20 GeV in \cite{Li:2021gsx}.

%%%%%%%%%%%%%%%%%%%%%%%%%%%%%%%%%%%%%%%%%%%%%%%%%%%%%%%%%%%%%%%

\section{BARYON STATIC PROPERTIES}

\subsection{Proton Mass and Pole Residue}

We begin with the two-point correlation function 
\cite{Ioffe:1981kw,Reinders:1984sr,Colangelo:2000dp}
\begin{equation}
\Pi(k)=i\int d^4x e^{ik\cdot x}\langle 0|T[\eta(x)\bar\eta(0)]|0\rangle
=A(k^2)k\!\!\!/+B(k^2)\;,\label{corn}
\end{equation}
where $\eta(x)=\epsilon^{abc}[u^{aT}(x)C\gamma^\mu u^b(x)]\gamma_5\gamma_\mu d^c(x)$ is a 
baryonic interpolating current, and $A(k^2)$ ($B(k^2)$) denotes the chiral-even (chiral-odd) 
part of the correlator. Equation~(\ref{corn}) describes a modified proton propagator at 
hadron level, 
\begin{equation}
    \Pi(k)=\lambda_P^2\frac{k\!\!\!/+m_P}{m_P^2-k^2}
%+\lambda_-^2\frac{k\!\!\!/-m_-}{m_-^2-k^2}
    +...\;,
\end{equation}
with the proton mass $m_P$ and the pole residue $\lambda_P$. Hence, the functions $A(k^2)$ 
and $B(k^2)$ correspond to
\begin{equation}
    A(k^2)=\frac{\lambda_P^2}{M_P^2-k^2},\quad B(k^2)=\frac{\lambda_P^2M_P}{M_P^2-k^2}.
    \label{ab}
\end{equation}

The OPEs for the functions $A(k^2)$ and $B(k^2)$ are given by \cite{Ioffe:1981kw}
\begin{equation}
    \begin{split}
        &\Pi_A(k^2)=\frac{k^4}{64\pi^4}\ln\left(\frac{-k^2}{\mu^2}\right)
        +\frac{\langle\alpha_sG^2\rangle}{32\pi^3}\ln\left(\frac{-k^2}{\mu^2}\right)
        -\frac{2}{3k^2} \kappa\langle\bar{q}q \rangle^2\\
        &\Pi_B(k^2)=-\left(\frac{k^2}{4\pi^2}\langle\bar{q}q\rangle +\frac{1}{8\pi^2}
        \langle\bar{q}g\sigma\cdot Gq\rangle\right)\ln\left(\frac{-k^2}{\mu^2}\right),
    \end{split}
\end{equation}
which contain dimension-even and dimension-odd condensates, respectively. It is noticed that
an additional quark-gluon condensate $\langle\bar{q}g\sigma\cdot Gq\rangle$ appears,
compared to the $\rho$ meson case. We derive the dispersion relations straightforwardly 
\begin{equation}
    \begin{split}
        &\frac{1}{\pi}\int_0^R ds\frac{\rho_A(s)}{s-k^2}
        =\frac{1}{\pi }\int_0^R ds\frac{\text{Im}\Pi_{A}(s)}{s-k^2}
        -\frac{2}{3k^2} \kappa\langle \bar{q}q \rangle^2,\\
        &\frac{1}{\pi}\int_0^R ds\frac{\rho_B(s)}{s-k^2}=
        \frac{1}{\pi}\int_0^R ds\frac{\text{Im}\Pi_B(s)}{s-k^2},
    \end{split}\label{eq:sumrule3}
\end{equation}
with the spectral densities $\rho_A(s)\equiv\text{Im}A(s)/\pi$ and 
$\rho_B(s)\equiv\text{Im}B(s)/\pi$, and
\begin{equation}
    \begin{split}
        &\text{Im}\Pi_{A}(s)=\pi\left(\frac{s^2}{64\pi^4}
        +\frac{\langle\alpha_sG^2\rangle}{32\pi^3}\right),\\
        &\text{Im}\Pi_B(s)=-\pi\left(\frac{s}{4\pi^2}\langle\bar{q}q\rangle 
        +\frac{1}{8\pi^2}\langle\bar{q}g\sigma\cdot Gq\rangle\right).
    \end{split}
\end{equation}

We define the subtracted spectral densities,
\begin{equation}
    \begin{split}
        &\Delta\rho_A(s,\Lambda)=\rho_A(s)-\frac{s^2}{64\pi^4}[1-\exp(-s/\Lambda)]
        -\frac{\langle\alpha_sG^2\rangle}{32\pi^3}[1-\exp(-s^2/\Lambda^2)],\\
        &\Delta\rho_B(s,\Lambda)=\rho_B(s)+\frac{s}{4\pi^2}
        \langle\bar{q}q\rangle[1-\exp(-s/\Lambda)] 
        +\frac{1}{8\pi^2}\langle\bar{q}g\sigma\cdot Gq\rangle[1-\exp(-s^2/\Lambda^2)],
    \end{split}\label{psub}
\end{equation}
where the subtraction factor $1-\exp(-s^2/\Lambda^2)$ for the last pieces goes down like 
$s^2$ as $s\rightarrow{0}$. It is introduced in view that the condensates are
constant, so stronger suppression is necessary for alleviating their impact on the spectral
densities $\rho_{A,B}(s)$ in the small $s$ region. The variable changes $x=k^2/\Lambda$ and 
$y=s/\Lambda$ lead to the dispersion relations for the subtracted spectral densities 
\begin{equation}
    \begin{split}
        &\int_0^\infty dy\frac{\Delta\rho_A(y)}{x-y}=\frac{1}{64\pi^4}\int_0^\infty dy
\frac{y^2e^{-y}}{x-y}+\frac{\langle\alpha_sG^2\rangle}{32\pi^3\Lambda^2}
\int_0^\infty dy\frac{e^{-y^2}}{x-y}+\frac{2}{3x\Lambda^3} \kappa\langle \bar{q}q \rangle^2,\\
        &\int_0^\infty dy\frac{\Delta\rho_B(y)}{x-y}=
        -\frac{\langle\bar{q}q\rangle}{4\pi^2\Lambda^{3/2}}\int_0^\infty dy\frac{ye^{-y}}{x-y}
-\frac{\langle\bar{q}g\sigma\cdot Gq\rangle}{8\pi^2\Lambda^{5/2}}\int_0^\infty dy\frac{e^{-y^2}}{x-y},
    \end{split}\label{eq:sumrule4}
\end{equation}
where the arbitrary scale $\Lambda$, substituted for the arbitrary cutoff $R$ in 
Eq.~(\ref{eq:sumrule3}), has migrated into the condensate terms to make them dimensionless.
Note that the dimensionless unknown functions $\Delta\rho_A(y)$ and $\Delta\rho_B(y)$ stand for
$\Delta\rho_A(s,\Lambda)/\Lambda^2$ and $\Delta\rho_B(s,\Lambda)/\Lambda^{5/2}$, respectively.

%hadrons with positive and negative parities. 

We then solve for $\Delta\rho_{A,B}(y)$ from Eq.~(\ref{eq:sumrule4}) with the inputs
\begin{equation}
\begin{split}
    &\langle\bar{q}q\rangle=-(0.215)^3\;\text{GeV}^3,\quad
    \langle \alpha_sG^2\rangle=0.028\;\text{GeV}^2,\quad
    \langle\bar{q}g\sigma\cdot Gq\rangle=m_0^2\langle\bar{q}q\rangle,\\
    &m_0^2=1.05\;\text{GeV}^2,\quad\kappa=2.
    \end{split}\label{newpND}
\end{equation}
The quark, gluon and four-quark condensates, being universal, take the same values as in 
Eq.~(\ref{cond}). The additional parameter $m_0^2$ associated with the quark-gluon 
condensate $\langle\bar{q}g\sigma\cdot Gq\rangle$ is set to a value around the upper 
bound of the range quoted in \cite{Ioffe:1981kw}. However, difficulty is encountered 
immediately; no solutions, which are stable against the variation of the polynomial 
number $N$, are found in a baryon case, indicating nontrivial disparity between the 
dispersive analyses on meson and baryon systems. 

The reason why the procedure outlined in the previous section does not work here will be 
elaborated. There are two corrections with different powers of $1/x$ in the OPE for the 
$\rho$ meson case on the right-hand side of Eq.~(\ref{sumrulerho_changed}), i.e., the 
$1/x^2$ piece proportional to the quark and gluon condensates and the $1/x^3$ piece 
proportional to the four-quark condensate. Their effects on the subtracted spectral density 
$\Delta\rho(y)$ are opposite; the combination of the quark and gluon condensates has a 
negative coefficient, and the inverse matrix elements $(M^{-1})_{n2}$ are negative, such 
that their products contribute to $\Delta\rho(y)$ constructively. The four-quark condensate 
has a negative coefficient, and contributes destructively, after being multiplied by the 
positive inverse matrix elements $(M^{-1})_{n3}$. We have $|(M^{-1})_{n2}|>|(M^{-1})_{n3}|$ 
for small $n$, so the $1/x^2$ effect dominates. As $n$ increases, $|(M^{-1})_{n3}|$ rises 
much faster than $|(M^{-1})_{n2}|$, and the $1/x^3$ term becomes important gradually. When 
the two pieces are comparable to each other, results for $\Delta\rho(y)$ are stabilized. 
This observation is in line with the one in \cite{Li:2020ejs} that both the $1/x^2$ and 
$1/x^3$ contributions are necessary for the emergence of the $\rho$ resonance. In the 
proton case there is only one power correction, the $1/x$ term, in the first formula of 
Eq.~(\ref{eq:sumrule4}). Therefore, its effect cannot be moderated by others, and no 
stable solutions in $N$ for the subtracted spectral density exist. 

\begin{figure}
    \centering
    \includegraphics[width=0.75\linewidth]{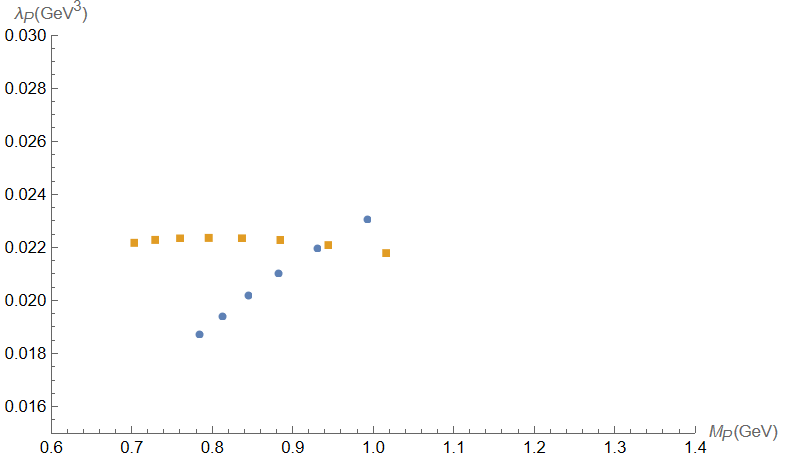}\quad    
    \includegraphics[width=0.75\linewidth]{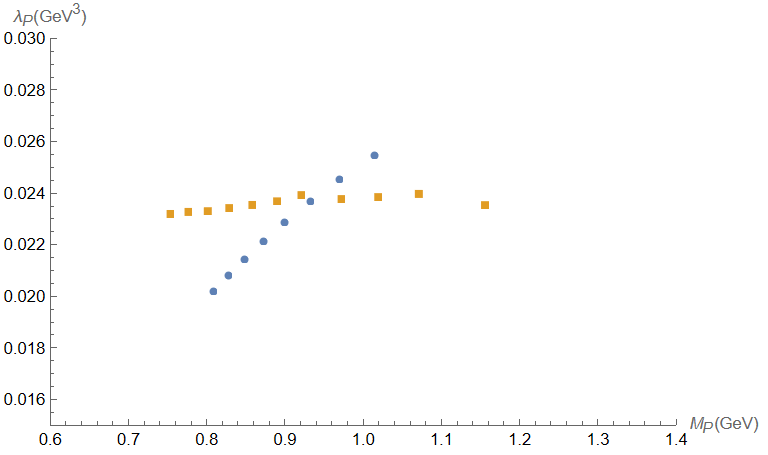}\quad
    \includegraphics[width=0.75\linewidth]{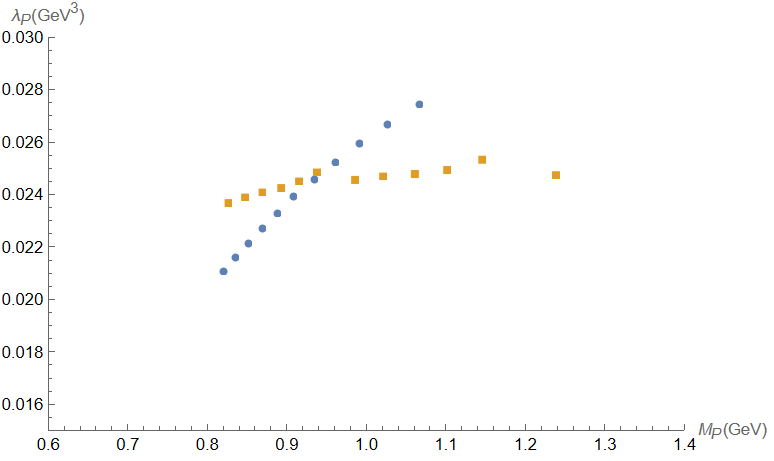}    
    \caption{Proton masses $M_P$ and pole residues $\lambda_P$ derived from the chiral-even 
    dispersion relation (dots) and the chiral-odd dispersion relation (squares) for $\Lambda=2$ 
GeV$^2$, 3 GeV$^2$ and 4 GeV$^2$ from top to bottom.}\label{fig:NLa2}
\end{figure}

We switch to a new strategy. Given a $\Lambda$ value, we obtain a solution for 
$\rho_{A}(s)$ for a polynomial number $N$, extract the proton mass squared 
$M_P^2$ from the location of the first peak in $s$, and compute the pole residue squared 
$\lambda_P^2$ according to Eq.~(\ref{dec}), i.e., the area under the first peak of 
$\Delta\rho_{A}(s,\Lambda)$. The spectral density $\rho_A(s)$ corresponding to the 
considered $\Lambda$ and $N$ must respect the requirement of positivity, of course. The 
resultant set of $\{M_P,\lambda_P\}$ formed by scanning the number $N$ represents a curve on 
the $M_P$-$\lambda_P$ plane as shown in Fig.~\ref{fig:NLa2}, where the discrete points 
correspond to various integers $N$. The solutions for 
$\rho_B(s)$ produce another curve on the $M_P$-$\lambda_P$ plane. Note that 
Eq.~(\ref{dec}) for $\Delta\rho_B(s,\Lambda)$ gives rise to $\lambda_P^2M_P$ according to 
Eq.~(\ref{ab}). The above two curves intersect each other, specifying unique values of 
$M_P$ and $\lambda_P$, which are regarded as the solution for the given $\Lambda$. The 
intersections of the curves from $\rho_A(s)$ and $\rho_B(s)$ 
for $\Lambda=2$ GeV$^2$, 3 GeV$^2$ and 4 GeV$^2$ are evident in Fig.~\ref{fig:NLa2}. We 
remark that the intersection, i.e., the solution is labeled by a pair of non-integer 
$N$'s, strictly speaking, one from $\rho_A(s)$ and the other from 
$\rho_B(s)$. It differs from the $\rho$ meson case, where a solution is 
labeled by a single $N$. 

\begin{figure}[h!]
    \centering
    \includegraphics[width=0.45\linewidth]{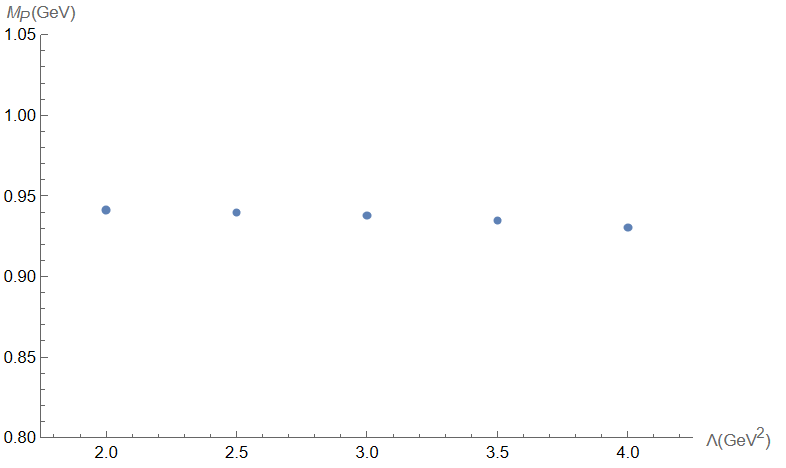}\quad    
    \includegraphics[width=0.45\linewidth]{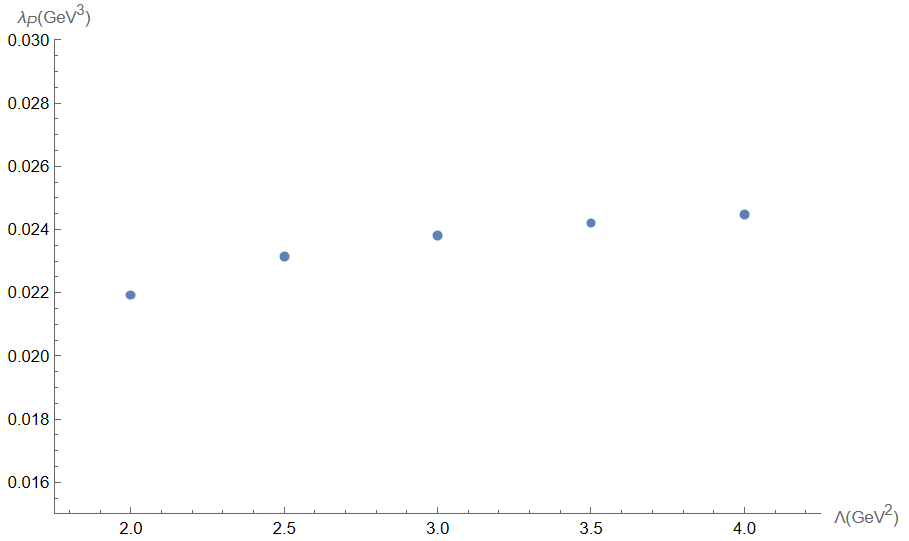}
    \caption{Dependencies of the proton mass $M_P$ and pole residue $\lambda_P$
    on $\Lambda$.}\label{fig:NL}
\end{figure}

It is encouraging that the intersections of the two curves from the chiral-even and 
chiral-odd dispersion relations about $M_P$=0.93 GeV and $\lambda_P=0.023$ GeV$^3$ are 
insensitive to the variation of the arbitrary scale $\Lambda$. That is, a stability window 
in $\Lambda$ does exist for the proton static properties, which is indicated by the 
plateaus of $M_P$ and $\lambda_P$ in the interval $[2\;\text{GeV}^2,4\;\text{GeV}^2]$ of 
$\Lambda$ in Fig.~\ref{fig:NL}. We conclude the proton mass and pole residue 
\begin{eqnarray}
M_P=(0.936\pm 0.005)\; {\rm GeV},\;\;\;\;\lambda_P= (0.0232\pm 0.0013) {\rm GeV}^3,
\end{eqnarray} 
where the former matches the data in \cite{PDG}, and the latter is lower than 
$\lambda_P= 0.0346$ GeV$^3$ in \cite{Ioffe:1981kw}. The pole 
residue, changing by about 10\% within the stability window, discloses a slight dependence 
on $\Lambda$. The similar trend of the $\rho$ meson decay constant $f_\rho$ derived from 
Eq.~(\ref{dec}) has been also observed. This $\Lambda$ dependence is not unexpected 
\cite{Li:2021gsx}, since Eq.~(\ref{dec}) is not a rigorous definition. For instance, 
truncating the integration variable $s$ at the first valley of the subtracted spectral 
density is somewhat discretionary. In summary, imposing the dispersive constraints from the 
chiral-even and chiral-odd parts of the correlator at the same time enforces the balance 
among all the available condensate contributions, such that a physical solution can be 
established. 

\begin{figure}[h]
    \centering
    \includegraphics[width=0.75\linewidth]{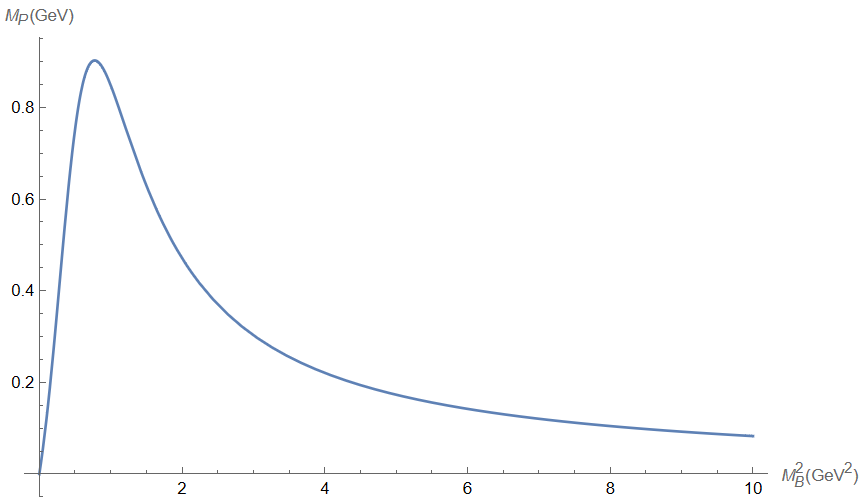}
    \caption{Dependence of the proton mass $M_P$ on the Borel mass $M_B$ in conventional 
    QCD sum rules with the inputs in Eq.~(\ref{newpND}).}
    \label{fig:1}
\end{figure}

At last, we compare our approach to the conventional treatment of QCD sum rules.
One applies the Borel transformation to the chiral-even and chiral-odd dispersion relations, 
and then takes their ratio to cancel the pole residue, 
yielding the proton mass
\cite{Ioffe:1981kw}
\begin{equation}
    M_P(M_B^2)=\frac{-\frac{\langle\bar{q}q\rangle}{4\pi^2}M_B^4
    -\frac{\langle\bar{q}g\sigma\cdot Gq\rangle}{8\pi^2}M_B^2}{\frac{M_B^6}{32\pi^4}
    +\frac{2}{3} \kappa\langle\bar{q}q\rangle^2},
    \label{eq:mass}
\end{equation}
with the Borel mass $M_B$. In principle, one ought to look for a stability window in $M_B$,
within which the proton mass $M_P$ remains constant roughly; a physical quantity should not 
depend on the arbitrary Borel mass. However, it is apparent that $M_P$ changes with $M_B$
dramatically, as revealed in Fig.~\ref{fig:1} with the inputs in Eq.~(\ref{newpND}). The 
author of \cite{Ioffe:1981kw} had to assume the accidental equality of the Borel mass 
and the proton mass in order to get an appropriate result. Alternatively, one selects a window 
in the Borel mass by allocating certain weights to the OPE continuum contribution and 
condensates \cite{Azizi:2016hbr,MarquesL:2018fph}. As a contrast, our predictions for the 
proton mass and pole residue manifest satisfactory stability against the variation of  
$\Lambda$ in Fig.~\ref{fig:NL}. That is, there is little ambiguity in making predictions in 
our formalism.

%%%%%%%%%%%%%%%%%%%%%%%%%%%%%%%%%%%%%%%%%%%%%%%%%%%%%%%%%%%%%%%%%

\subsection{$\Delta(1232)$ Mass and Pole Residue}

%(if the anomalous dimensions and excited states' contributions are ignored)

The generalization to the investigation of a $\Delta(1232)$ state in the decuplet is 
straightforward. We just need to consider the $g^{\mu\nu}$ structure in the corresponding 
two-point correlation 
function 
\begin{equation}
\Pi^{\mu\nu}(k^2)=i\int d^4x e^{ik\cdot x}\langle 0|T\{\eta^\mu(x)\bar\eta^\nu(0)\}|0\rangle
=-g^{\mu\nu}[A(k^2)k\!\!\!/+B(k^2)]+\cdots\;,
\end{equation}
with the current $\eta^\mu(x)=\epsilon^{abc}[u^{aT}(x)C\gamma^\mu u^b(x)]u^c(x)$, 
the chiral-even part $A(k^2)$ and the chiral-odd part $B(k^2)$. We adopt 
Ioffe's results for the OPEs \cite{Ioffe:1981kw}
\begin{equation}
    \begin{split}
        \Pi_{A}(k^2)&=\frac{k^4}{160\pi^4} \ln\left(\frac{-k^2}{\mu^2}\right)
        -\frac{4}{3k^2} \kappa\langle\bar{q}q\rangle^2,\\
        \Pi_{B}(k^2)&=-\left[\frac{k^2\langle\bar{q}q\rangle}{3\pi^2} 
        -\frac{\langle\bar{q}g\sigma Gq\rangle}{24\pi^2}\right]\ln\left(\frac{-k^2}{\mu^2}\right).
        %\\
       %\Pi^{(\text{pert}')}_{\text{Ioffe}}(k^2)&=\frac{k^4}{160\pi^4} \ln\left(\frac{-k^2}{\mu^2}\right).
    \end{split}\label{iof}
\end{equation}
Lee's expressions \cite{Lee:1997ix} agree with Ioffe's except the coefficient of 
the quark-gluon condensate,  
\begin{equation}
    \begin{split}
        \Pi'_{A}(k^2)&=\frac{k^4}{160\pi^4} \ln\left(\frac{-k^2}{\mu^2}\right)
        -\frac{4}{3k^2} \kappa\langle\bar{q}q\rangle^2,\\
        \Pi'_{B}(k^2)&=-\left[\frac{k^2\langle\bar{q}q\rangle}{3\pi^2} 
        -\frac{\langle\bar{q}g\sigma Gq\rangle}{6\pi^2}\right]\ln\left(\frac{-k^2}{\mu^2}\right).
        %\\
       %\Pi^{(\text{pert}')}_{\text{Frank}}(k^2)&=\frac{k^4}{160\pi^4} \ln\left(\frac{-k^2}{\mu^2}\right).
    \end{split}\label{lee}
\end{equation}
We repeat the OPE calculation, and find that Eq.~(\ref{iof}) is favored. In fact, we 
cannot construct physical solutions to the spectral densities from Eq.~(\ref{lee}) under 
the requirement of positivity, because the quark-gluon condensate gives a sizable 
destructive contribution.

We define the subtracted spectral densities
\begin{equation}
    \begin{split}
        &\Delta\rho_{A}(s,\Lambda)=\rho_{A}(s)-\frac{s^2}{160\pi^4}[1-\exp(-s/\Lambda)],\\
        &\Delta\rho_{B}(s,\Lambda)=\rho_{B}(s)
        +\frac{s}{3\pi^2}\langle\bar{q}q\rangle[1-\exp(-s^2/\Lambda^2)] 
        -\frac{1}{24\pi^2}\langle\bar{q}g\sigma\cdot Gq\rangle
        [1-\exp(-s^3/\Lambda^3)].\\
    \end{split}
\end{equation}
The spectral densities associated with a heavier $\Delta$ are expected to have
softer behaviors near the origin $s=0$. Hence, we employ the subtraction factors
that fall faster as $s\to 0$ than in the proton case. For example, the constant
quark-gluon condensate is suppressed by $1-\exp(-s^3/\Lambda^3)\propto s^3$ to
lessen its impact on the spectral density $\rho_{B}(s)$ at low $s$.
The dispersion relations for the subtracted spectral densities read
\begin{equation}
    \begin{split}
        &\int_0^\infty dy\frac{\Delta\rho_{A}(y)}{x-y}
        =\frac{1}{160\pi^4}\int_0^\infty dy\frac{y^2e^{-y}}{x-y}
        +\frac{4}{3x\Lambda^3} \kappa\langle \bar{q}q \rangle^2,\\
        &\int_0^\infty dy\frac{\Delta\rho_{B}(y)}{x-y}=
        -\frac{\langle\bar{q}q\rangle}{3\pi^2\Lambda^{3/2}}
        \int_0^\infty dy\frac{ye^{-y^2}}{x-y}
        +\frac{\langle\bar{q}g\sigma\cdot Gq\rangle}{24\pi^2\Lambda^{5/2}}
        \int_0^\infty dy\frac{e^{-y^3}}{x-y}.\\
    \end{split}\label{eq:IOFFESUM}
\end{equation}

\begin{figure}[t!]
    \centering
    \includegraphics[width=0.75\linewidth]{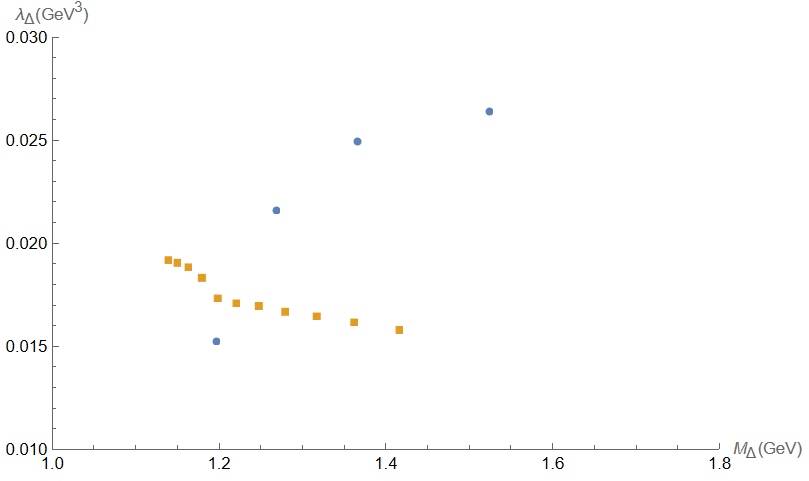}\quad    
    \includegraphics[width=0.75\linewidth]{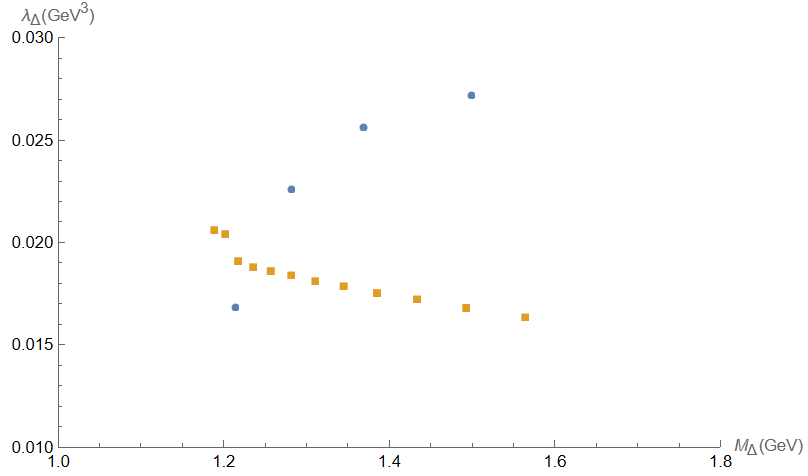}\quad
    \includegraphics[width=0.75\linewidth]{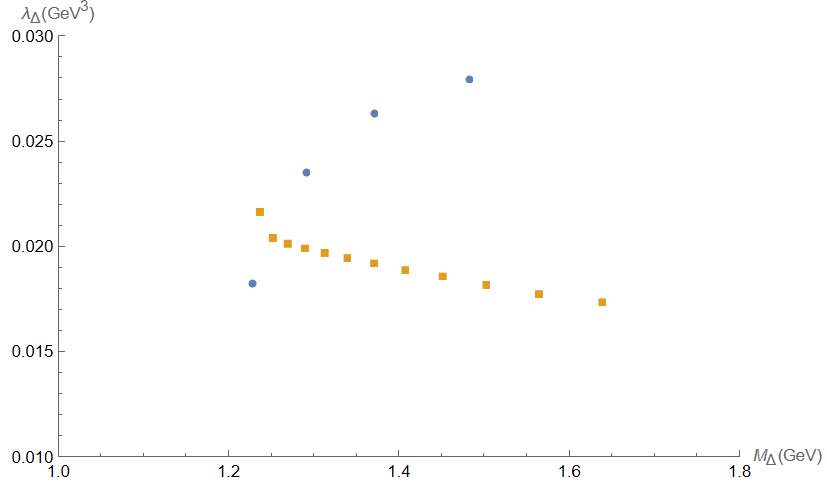}    
    \caption{$\Delta(1232)$ masses $M_\Delta$ and pole residues $\lambda_\Delta$ derived 
    from the chiral-even dispersion relation (dots) and the chiral-odd dispersion relation 
    (squares) for $\Lambda=1.8$ GeV$^2$, 2.0 GeV$^2$ and 2.2 GeV$^2$ from top to bottom.}
    \label{fig:DLa3}
\end{figure}

Inputting the same parameters as in Eq.~(\ref{newpND}) for the proton case, we present 
in Fig.~\ref{fig:DLa3} the curves from $\rho_A(s)$ and $\rho_B(s)$ on the 
$M_\Delta$-$\lambda_\Delta$ plane for $\Lambda=1.8$ GeV$^2$, 2.0 GeV$^2$ and 2.2 GeV$^2$.
Note that we infer a resonance mass through the subtracted spectral density 
$\Delta\rho_B(s,\Lambda)$ actually, because its peaks are more prominent than those 
of $\rho_B(s)$ after the subtraction in Eq.~(\ref{psub}). Similar situation has occurred 
in the dispersive analyses on glueball masses \cite{Li:2021gsx}. This alternative 
affects only a little bit the results for the $\Delta(1232)$ baryon. Again, the two curves 
in Fig.~\ref{fig:DLa3} intersect each other, designating unique values of $M_\Delta$ and 
$\lambda_\Delta$ for each $\Lambda$. The dependencies of $M_\Delta$ and $\lambda_\Delta$
on $\Lambda$ are similar to those in Fig.~\ref{fig:NL} with acceptable stability, so they 
are not displayed for simplicity. The difference is that the stability window in $\Lambda$ is 
narrower in the $\Delta(1232)$ case than in the proton one. We deduce the $\Delta(1232)$ mass 
and pole residue from Fig.~\ref{fig:DLa3}
\begin{eqnarray}
M_\Delta=(1.238\pm 0.018)\;{\rm GeV}, \;\;\;\;
\lambda_\Delta=(0.019\pm 0.002)\;{\rm GeV}^3,
\end{eqnarray}
where the former is consistent with the data in \cite{PDG}, and the latter is lower than 
but of the same order as $\lambda_\Delta=0.05$ GeV$^3$ in \cite{Ioffe:1981kw}.

After completing all the formulas, we are ready to explain how to choose the 
inputs for the condensates involved in this work. We fix the factorization violation 
parameter for the four-quark condensate to $\kappa=2$ for an illustration. The OPEs 
associated with a $\Delta(1232)$ baryon are insensitive to the quark-gluon condensate 
owing to the tiny coefficient $1/(24\pi^2)$ and independent of the gluon condensate. The 
quark condensate is the most crucial for $\Delta(1232)$ properties, and thus set first. 
Secondly, the quark-gluon condensate does not contribute in the $\rho$ meson case, so 
the gluon condensate is tuned to match the $\rho$ meson data without altering the 
$\Delta(1232)$ outcomes. Finally, the parameter $m_0^2$ in the parametrization of the 
quark-gluon condensate is adjusted to account for proton observables with negligible 
modification of the $\Delta(1232)$ results. All the values for the above condensates are 
within their allowed ranges known in the literature. Therefore, the point of our analysis 
resides in the possibility of establishing unambiguous and stable solutions to the 
dispersion relations for meson and baryon systems with universal and reasonable condensate 
inputs.

\section{CONCLUSION}

We have extended the dispersive studies on meson static properties to those of light 
baryons. The challenge arises from the structure of a baryonic correlation function distinct
from a mesonic one; the former is split into a chiral-even part and a chiral-odd part,
such that the number of power corrections for each chirality is insufficient, and a physical 
solution cannot be identified from each dispersion relation. To surmount the difficulty, we 
implemented the chiral-even and chiral-odd dispersive constraints simultaneously, and found 
the common solutions for the baryon masses and pole residues. It has been corroborated that 
the obtained solutions are stable against the variation of the arbitrary scale $\Lambda$, 
which traces back to the ultraviolet cutoff for the dispersive integrals. We have also 
commented on the conventional handling of QCD sum rules; no stability window in the 
Borel mass appears in the proton mass sum rule actually. Our approach is thus superior to 
the conventional one. Motivated by the universality of condensates, we have 
demonstrated that the same set of quark and gluon condensates can accommodate the masses 
of a $\rho$ meson, a proton and a $\Delta(1232)$ baryon.

This work will stimulate broad applications of our nonperturbative analytical framework 
not only to mesonic but baryonic quantities. We will apply it to the determination of more 
complicated baryon light-cone distribution amplitudes, which are the 
essential nonperturbative inputs to the factorization theorem for high-energy baryonic 
processes and heavy baryon decays, but remain quite uncertain \cite{Han:2024kgz}. Besides, 
a global fit of various quark and gluon condensates to relevant data can be done, and fit 
results, based on their universality, can be employed to make predictions for other 
hadronic observables. The agreement of fit results with those evaluated in lattice QCD will 
offer a convincing support to our approach.

%%%%%%%%%%%%%%%%%%%%%%%%%%%%%%%%%%%%%%%%%%%%%%%%%%%%%%%%%
%%%%%%%%%%%%%%%%%%%%%%%%%%%%%%%%%%%%%%%%%%%%%%%%%%%%%%%%%
\section*{Acknowledgement}

We thank W. Chen, H. Mutuk and A.S. Xiong for fruitful discussions. This work was supported 
by the National Science Foundation of China (NSFC) under Grant No. 12475095 and by 
National Science and Technology Council of the Republic of China under Grant 
No. NSTC-113-2112-M-001-024-MY3.

%%%%%%%%%%%%%%%%%%%%%%%%%%%%%%%%%%%%%%%%%%%%%%%%%%%%%%%%
%%%%%%%%%%%%%%%%%%%%%%%%%%%%%%%%%%%%%%%%%%%%%%%%%%%%%%%%

\end{document}